# Haina Storage: A Decentralized Secure Storage Framework Based on Improved Blockchain Structure

Zijian Zhou[1, 3, 4], Caimei Wang[2, *], Xiaoheng Deng[3, 4], Jianhao Lu[2], Qilue Wen[2], Chen Zhang[2], and Hong Li[2]

[1] School of Software, Xinjiang University, Urumqi 830046, China
[2] School of Artificial Intelligence and Big Data, Hefei University, Hefei 230601, China
[3] School of Electronic Information, Central South University, Changsha 410083, China
[4] Shenzhen Research Institute, Central South University, Shenzhen, 518000, China
[*] Corresponding author. Email: wangcm@hfuu.edu.cn

**Abstract.** Although the decentralized storage technology based on the blockchain can effectively realize secure data storage on cloud services. However, there are still some problems in the existing schemes, such as low storage capacity and low efficiency. To address related issues, we propose a novel decentralized storage framework, which mainly includes four aspects: (1) we proposed a Bi-direction Circular Linked Chain Structure (BCLCS), which improves data's storage capacity and applicability in decentralized storage. (2) A Proof of Resources (PoR) decision model is proposed. By introducing the network environment as an essential evaluation parameter of storage right decision, the energy and time consumption of decision-making are reduced, and the fairness of decision-making is improved. (3) A chain structure dynamic locking mechanism (CSDLM) is designed to realize anti-traverse and access control. (4) A Bi-directional data Access Mechanism (BDAM) is proposed, which improves the efficiency of data access and acquisition in decentralized storage mode. The experimental results show that the framework has significantly improved the shortcomings of the current decentralized storage.

## 1. Introduction

With the advent of big data, migrating data to the cloud is essential to alleviate the shortage of local storage space. Data security is the biggest challenge of cloud storage today [1]. The core issue of ample data research is ensuring safe data storage in the cloud effectively. Currently, the primary data storage mode is centralized. That is, data is stored in a reliable data center. In this mode, data encryption and access control are the most commonly used methods to protect data security [2-4]. However, this centralized storage mode has a single point of failure, an unreliable data center, and other problems.

The decentralized approach can effectively solve the above problems. Blockchain is one of the typical decentralized technologies proposed by Nakamoto [8]. The blockchain consists of different blocks connected into a chain in the chronological order in which they were created. Each block stores a certain amount of data. It has been applied in many aspects, such as justice, medical treatment [6-7], etc. However, all nodes in the decentralized network are required to store all blocks, which causes capacity wasting. More redundancy is demanded with data expansion, so blockchain isn't suitable anymore.

IPFS is another decentralized storage scheme that replaces the HTTP protocol [5]. Combined IPFS and other approaches like deep learning have contributed to fault detection, the Internet of Things, and other fields [34-35]. IPFS is more likely a public sharing resources network for storage solutions than a personal cloud disk. Due to the public data acquisition approach, individuals and private data storage uploaded to IPFS are still skeptical.

Many studies [9-16] try to reduce the storage capacity and improve the security of data. They can be divided into on-chain and off-chain means. In the on-chain method, all varieties of data replace the transaction data. This method secures the safety of data but ignores capacity. After that, the off-chain scheme appeared [27]. For example, references [11-16] use an off-chain model, as Figure 1 presented for file storage. Blockchain acts as a mapping role in this model and stores the relationship between file information and storage location as a block in the blockchain.

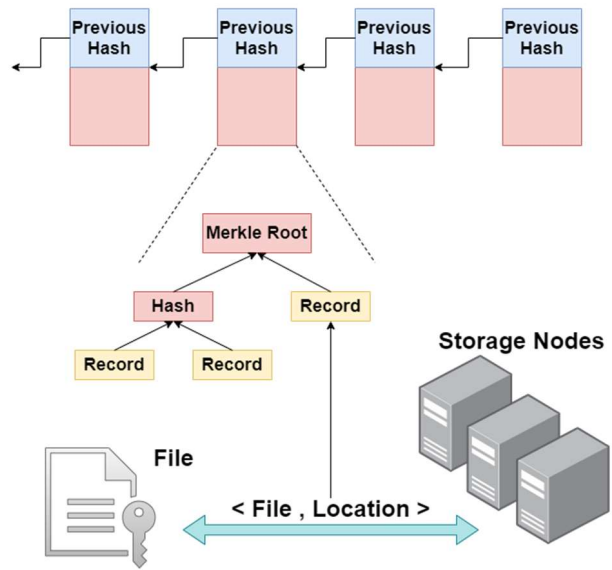

**Fig. 1.** Off-chain Schematic

The off-chain method reduced the needed storage space by detaching the actual data from the block, thereby decreasing the need for storage capacity in the blockchain. Storage capacity demand is indeed slowed down by this method. However, the off-chain model doesn't remove the blockchain, and the disadvantage of low capacity still exists. References [10, 17] try to rebuild the chain structure into tree and graph structures. However, the complex structure appends the cost of space to manage and store in addition.

Compared with the blockchain, the off-chain model and references [10, 17] have improved capacity and efficiency. They have in common that they improve by reconstructing the storage structure. Therefore, storage structure is essential for improving storage capacity and data acquisition efficiency. In this paper, we aim to explore the deeper factors underlying the current problem in data storage structure. And design practical solutions to achieve more massive data storage and higher data acquisition efficiency.

Besides, the number of storage nodes in a decentralized network is generally large, so it is crucial to design an appropriate decision mechanism to determine which storage nodes are used to store data. Unreasonable decision mechanisms often affect the overall security and performance of the framework. For example, an unfair decision mechanism may cause some nodes to store most of the data and thus degenerate into a centralized mode. If the decision mechanism requires all nodes to back up too much data, it may lead to colossal storage capacity requirements. Different mechanisms have been proposed to make reasonable decisions.

Three types of decision mechanisms are computing power-based, storage capacity-based, and family of byzantine fault tolerance—their typical mechanisms are shown in Table 1.

Table 1. The Typical Mechanism

| Type | Reference | Name |
| --- | --- | --- |
| Computing Power Based | [18] | Proof of Work (Pow) |
| | [19] | Proof of Stake (PoS) |
| | [20] | Delegated Proof of Work (DPoW) |
| | [21] | Delegated Proof of Stake (DPoS) |
| Storage Capacity Based | [22] | Proof of Capacity (PoC) |
| | [23] | Proof of Spacetime (PoSt) |
| Family of Byzantine Fault Tolerance | [25] | practical Byzantine Fault Tolerance (pBFT) |
| | [26] | Central Bank Digital Currency (CBDC) |
| | [27] | Stellar Consensus Protocol (SCP) |

The computing power base is easy to realize but involves serious energy waste, high time consumption, and lack of fairness. The storage capacity-based mechanism reduces energy waste, but it writes invalid data on the disk, and the utilization rate of the disk remains to be improved. The family of Byzantine Fault Tolerance [24] can expand, but it still needs to improve on the scale of nodes and anti-attack.

The data storage system should store data promptly. The network environments and storage node decision time are two significant aspects that impact the timeliness of a decentralized storage system. Network environments, such as network delay and data transfer rate, determine the data transfer time. And storage node decision time causes the delay in data transfer. Consequently, considering these two circumstances, a proper decision mechanism should be signed.

The above three mechanisms mainly focus on reducing the data transfer delay and preventing fault. Most of them lack consideration of the network environment. To make the storage decision more fair, secure, and efficient, we propose a storage-based decision mechanism called Proof of Resources (PoR), which combines the network status as the evaluation parameter. This part will be detailed in Section 4.3.

The main contributions are as follows:

(1) A **bi-directional circular L**inked **C**hain **S**tructure (BCLCS) is designed, which can solve the problems of decentralized storage, such as small storage capacity and low data storage matching.

(2) A **C**hain **S**tructure **D**ynamic **L**ocking **M**echanism (CSDLM) is designed to hide the correlation between data better and improve data security.

(3) a bi-directional data Access Mechanism (BDAM) is proposed to further improve the efficiency of the data access in a decentralized mode. Experiments show that this mechanism can improve data access efficiency by 38%.

(4) A **P**roof **o**f **R**esources (PoR) decision model is designed, in which the network environment is considered one of the most critical parameters because of its impact on storage. Compared with the existing decision models, our model reduces the computation force and improves the fairness of the strategy.

## 2. Preparation Work

### 2.1 Hash Pointer

Figure 2 shows the blockchain structure proposed by Nakamoto [8]. The structure is a one-way linear linked list that adds the previous block's hash value to the next. This inserted hash value is called a hash pointer that can uniquely identify the data content.

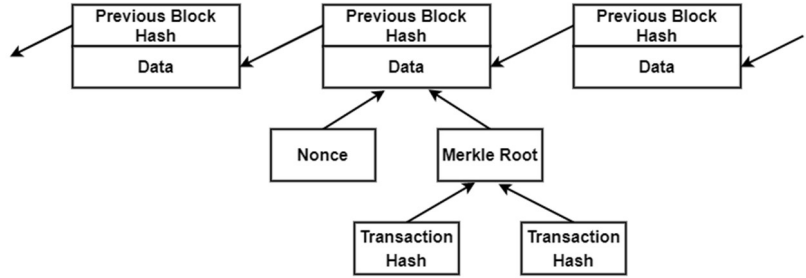

**Fig. 2.** Traditional Blockchain Data Storage Structure

The **B**i-directional **C**ircular **L**inked **C**hain **S**tructure (BCLCS) proposed in this paper divides the block into pointer domain and data domain and no longer relies on blockchain. It is developed from the basic blockchain structure. BCLCS discusses the deeper reasons for the blockchain's low storage capacity and efficiency. It effectively addresses the related problems.

### 2.2 Content Addressing

Because of the uniqueness of the hash value, IPFS identifies a file and blocks via its hash value. Files and blocks are stored in a decentralized network, using the hash value to search the storage location in the decentralized network to be defined as content addressing [5], which replaces the traditional URL addressing. This paper adopts content addressing to search a block in the decentralized network. Compared with the URL, the data obtained by content addressing will not change due to the location data change.

## 3. Composition of Framework

Before introducing the details of the framework, a general understanding of its composition will help in the overall comprehension of the framework. The framework is divided into the user and the decentralized network. The user, who can be recognized as a client, must use the SM4 algorithm to encrypt the data before uploading and then construct the ciphertext into a **B**i-direction **C**ircular **L**inked **C**hain **S**tructure (BCLCS). An inverse key storage mechanism has been proposed in our previous work [29], which is more applicable in a decentralized scheme. Therefore, we introduce it in this framework to improve the security of the encryption key.

To solve the ownership problem of a single data block in a decentralized network, we introduce the network environment as an essential evaluation parameter and propose a new decision model, Proof of Resources (PoR), based on the network environment and node storage capacity. Aiming to solve the problem of data security, the Chain Structure Dynamic Locking Mechanism (CSDLM) has been designed. To solve the problem of low data access efficiency, the **B**i-direction **D**ata

Acquisition Mechanism (BDAM), which imitates the characteristics of Deoxyribonucleic Acid (DNA) replication [30], is proposed, and the data access efficiency is effectively improved.

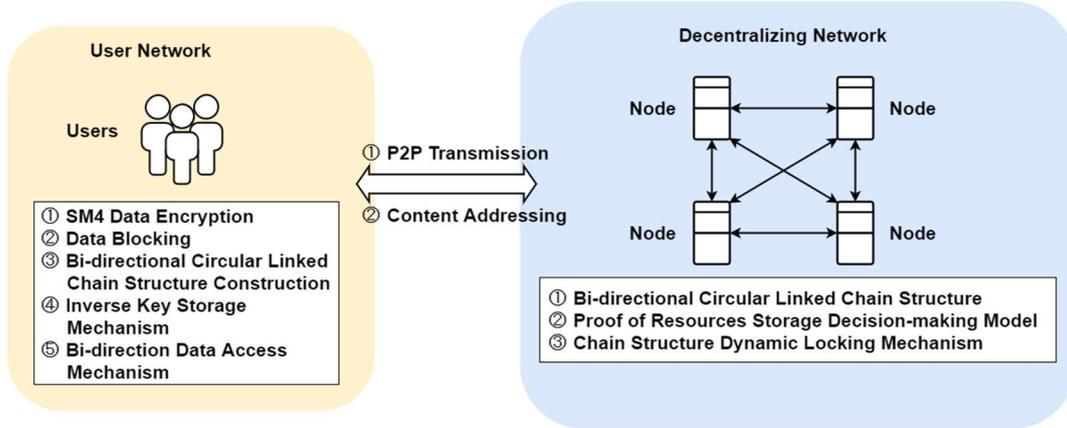

**Fig. 3.** Composition of Framework

## 4. Detail Design

In this section, we will introduce the **B**i-direction **C**ircular **L**inked **C**hain **S**tructure (BCLCS), the **C**hain **S**tructure **D**ynamic **L**ocking **M**echanism (CSDLM), and the **P**roof of **R**esources (PoR) in detail.

**4.1 Bi-direction Circular Linked Chain Structure (BCLCS)**

**4.1.1 Symbol Definition of BCLCS**

The symbols and meanings of BCLCS are shown in Table 2.

**Table 2.** Symbols of BCLCS

| Symbol | Description | Symbol | Description |
|---|---|---|---|
| $b$ | The data of blockchain is also the first dimension of blockchain. | $\varphi$ | The law in the definition of dependent. |
| $\varphi_1$ | The law in the definition of dependent. | $sizeof()$ | The capacity calculation algorithm. |
| $C$ | The chain structure of blockchain is also the second dimension of blockchain. | $O()$ | The time complexity or space complexity. |
| $A.x$ | The attribute $x$ of object $A$. | $t$ | The transaction data of the block. |
| $n$ | Quantity of storage nodes. | $m$ | Quantity of storage blocks. |
| $\Omega$ | The data set. | $previous\_hash$ | The hash pointer of the previous block. |
| $next\_hash$ | The hash pointer of the next block. | $current\_hash$ | The hash pointer of the current block. |
| $null$ | The empty content. | $H()$ | The hash algorithm. |

**4.1.2 Discussion on BCLCS**

In the introduction section, we stated that the data storage structure is the factor that impacts the problems of capacity and efficiency. This chapter will discuss and locate the deeper reason for low storage capacity and efficiency below.

Reviewing the blockchain structure, we found that it is a dependent structure. Different blocks continuously compose the one-direction chain, and some transaction data is stored in each block. This structure can be abstracted as a two-dimensional construction. The most minimal dimension is the block. The basic unit of a block is transaction data. Transaction data is denoted as $t$; each block of the blockchain is a finite set b, which is constituted by $t$. It is presented as $b = \{t_0, t_1, t_2, t_3, \ldots \ldots, t_n\}$. The second dimension is the chain $C$ which is an infinite set consisting of various blocks $b$. This form can be described as $C = \{b_0, b_1, b_2, b_3, \ldots \ldots\}$. The principle for blockchain to manufacture the chain that uses a law $\varphi$, which makes the elements $b$ of $C$ exist in the equation $b_i.previous\_hash = H(b_{i-1})$, function $H$ represents a hash algorithm.

**Definition 1:** In a set $\Omega$, the elements of $\Omega$ are dependent, if there is a law $\varphi$ at least which makes all the elements of $\Omega$ that can establish a relationship.

**Theorem 1:** The structure of blockchain is dependent.

**Proof 1:** We assume $\Omega$ is $C$, $C = \{b_0, b_1, b_2, b_3, \ldots \ldots\}$, and the law $\varphi$ as follows:
$$\varphi = \begin{cases} b_0.previous\_hash = null \\ b_i.previous\_hash = H(b_{i-1}) \end{cases}$$

Then all the elements of $C$ connect by the hash value which is called the hash pointer. Therefore, the blockchain $C$ is dependent.

Dependency leads to storage centralization, resulting in low storage capacity utilization. Storage centralization does not mean unachievable for a decentralized storage scheme. Instead, the dependent data can't be stored in a discrete form. In terms of blockchain, the chain must be stored in each node in aggregating form. The storage node couldn't verify the correctness and integrity if any block had been lost. Therefore, this kind of dependent attribute requires all nodes to store each block simultaneously, which means a gigantic redundancy has been caused that declines the storage capacity utilization and wastes the disk space. Blockchain is an infinite set, which implies that dependency will be prolonged with the expansion of blocks. However, the lengthening of dependency will extend the time cost when we try to search for some data in the chain by traversing the chain. Then, the low efficiency caused.

Presuming there are $n$ storage nodes and $m$ blocks. For blockchain, the demand for storage capacity is $n \times m \times sizeof(b)$, in the scale $O(nm)$. We can choose any node in the blockchain and traverse the chain to find a block, the scale of traversing is $O(m)$. Compared with these two measures, the reduction of $m$ will address the current major disadvantages to a certain degree. Unfortunately, $m$ is infinite, and the number of blocks is infinite.

### 4.1.3 Designation of BCLCS

After the discussion about dependency, the view of dependency supports a new solution. We have proof that blockchain is dependent, and state that dependency results in low storage capacity and low efficiency. Therefore, breaking the dependent property, in other words, transforming the storage data structure into an independent might solve the flaws successfully. We have believed that blockchain can abstract into two dimensions of construction, and the infinite blocks set $C$ is dependent. The reason caused the dependency is due to the law $\varphi$ which was created when the blockchain was designed. To rebuild the structure and perpetuate the integrity verification, we remove the law $\varphi$ to the first dimension $b$ and make infinite dependency into limited dependency. we redesign the law $\varphi$ to $\varphi_1$:

$$\varphi_1 = \begin{cases} t_i.previous\_hash = H(t_{i-1}.data) \\ t_i.hash = H(t_i.data) \\ t_i.next\_hash = H(t_{i+1}.data) \end{cases}$$

According to $\varphi_1$, as fig. 4. depicts, we propose the **B**i-direction **C**ircular **L**inked **C**hain **S**tructure (BCLCS). A data block is divided into two parts, namely, the pointer domain and the data domain. The pointer domain is used to store three hash pointer information, which is respectively used to point to the previous, current, and next blocks. These hash pointers are not only used to locate the adjacent blocks, but they are also undertaking the responsibility to verify the acracy and integrity of data. The mission of verifying each block has been transferred to the user after the block is uploaded. The data domain is used to carry the actual data. And, the finite set $b$ is built as a chain. BCLCS has the bi-direction feature because of the foundation of a finite set. In this paper, the block is not in the second dimension anymore, it turns to the first dimension.

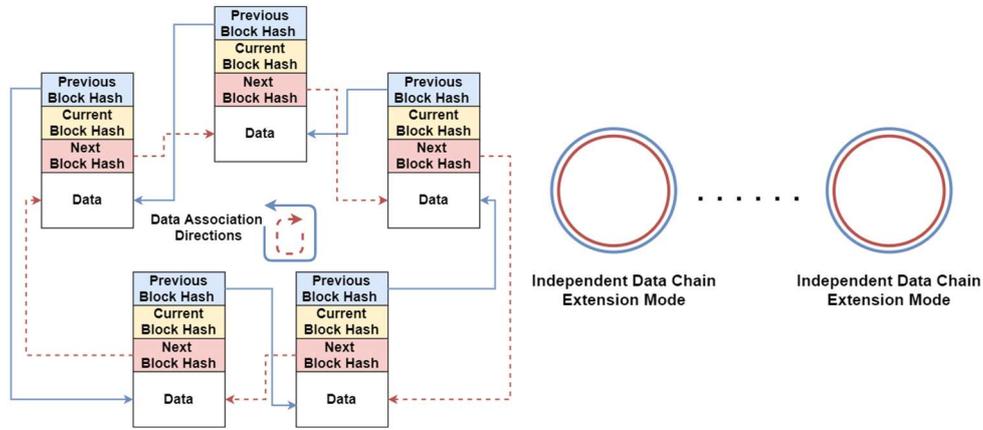

**Fig. 4.** The Bi-direction Circular Linked Chain Structure

After reconstruction, the second dimension $C$ is independent, which means that all the elements of $C$ stored in a node are unnecessary now and we can store each element dispersedly. Besides, at the original law $\varphi$, dependency is infinite which causes all of the original blocks must store centrally. This circumstance has improved in BCLCS, distributed storage is acceptable. Because we can search blocks through the pointer domain in the decentralized network, means we also do not require to store blocks in the same node. The distinctions are shown in Table 3.

**Table 3.** Comparison of Blockchain and BCLCS

| Object | Location of Block | Are All Blocks Store in A Node? | Are All Data Store in A Node for Blocks? | Block Searching Method |
|---|---|---|---|---|
| Blockchain | The 2nd. Dimension | Yes | Yes | Traverse the Chain in A Node |
| BCLCS | The 1st. Dimension | No | No | Search in Decentralized Network |

**4.2 Chain Structure Dynamic Locking Mechanism (CSDLM)**

**4.2.1 Symbol Definition of CSDLM**

The symbols and meanings of CSDLM are shown in Table 4.

Table 4. Symbols of CSDLM

| Symbol | Description | Symbol | Description |
|---|---|---|---|
| $RES$ | The beginning stages of transformation. | $Hash_{lock}$ | The original hash pointer. |
| $FIN$ | The final status of transform. | $Hash_{unlock}$ | The transformed hash pointer. |
| $f()$ | The transform function of CSDLM. | $Mask$ | The parameter of CSDLM. |
| $f^{-1}()$ | The reverse transform function of CSDLM. | $Meta\ File$ | The mapping information in this framework. |
| $p$ | The untransformed hash pointer. | $key$ | The encryption key for the encryption algorithm. |
| $p'$ | The transformed hash pointer. | $encrypt()$ | Using to encrypt data, SM4 is adopted in this paper. |
| $timestamp$ | Time indicator. | $decrypt()$ | Using to decrypt the file. |
| $EF$ | Encrypted file. | $extract()$ | Using to extract $Mask$. |
| $Update()$ | Using to update the node file. | $loc()$ | Using to locate the position of the $key$. |
| $\oplus$ | The XOR operation. | $HB_1$ | The hash pointer of the first block. |
| $\leftarrow$ | Presenting output to someplace. | $\xrightarrow{write}$ | The writing function. |

### 4.2.2 Discussion and Designation of CSDLM

When users try to store personal data or private data, it breaks the confidentiality of blocks if an attacker steals a block and obtains the whole chain by traversing the chain according to the pointer domain.

The law $\varphi$ makes such a situation possible. However, the framework would be paralysis, if the law were to vanish. We are aiming to solve this problem through a method that reserves the law $\varphi$ and achieves data protection at the same time. To avoid breaking the result of law $\varphi$, we only can rebuild the consequence, which is based on law $\varphi$. Figure 5 shows what we aim to do. The $RES$ is the status that presents what we have obtained now. $RES$ can be viewed as the result of $\varphi$, we can seek all blocks through $\varphi$. Function $f()$ transform the $RES$ to another circumstance $FIN$ which losses the ability to obtain all the blocks when the data are stored in the decentralized network. We could reacquire $RES$ by the reverse function $f^{-1}()$, if we wanted to recover the data we uploaded. To simplify the solution model, we make the equation $f() = f^{-1}()$, which is shown in figure 6.

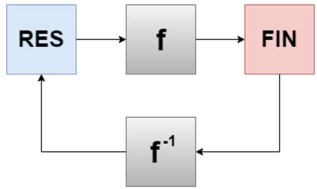
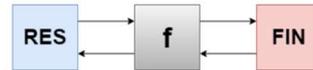

**Fig. 5.** The Solution Model  **Fig. 6.** The Reduced Solution Model

If we denote the hash pointer as $p$, then $p'$ is the answer we calculated by $f^{-1}()$. $p$ is a hash value, as only if $p \neq p'$, then the destinations are different of $p$ and $p'$. It means that we could not reserve the whole chain through $p'$, if we instead $p$ to $p'$.

According to the thought described above, we propose the Chain Structure Dynamic Locking Mechanism (CSDLM). This mechanism abstracts the data into two states, locking and unlocking, which are corresponding to $RES$ and $FIN$ respectively. Only in the unlocking state can the user obtain the complete data on the data chain. The data link status is shown in Fig. 7.

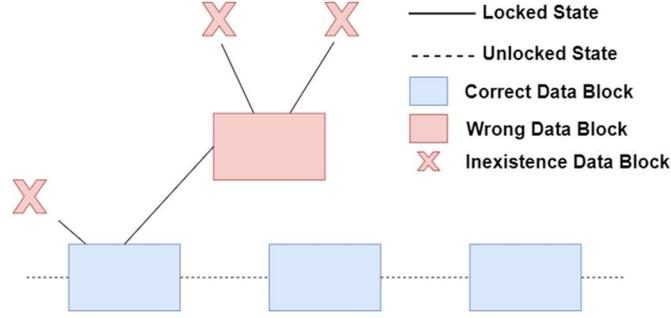

**Fig. 7.** Schematic Diagram of Lock and Unlock Status

The $f(\ ) = f^{-1}(\ )$ the form is consistent with the symmetric cipher. In order to further simplify the calculation and improve the efficiency, we define the function $f(\ )$ as XOR operation. At the same time, we randomly generate a parameter $Mask$ to satisfy the operation. We require $Mask \neq 0$, then the condition $p \neq p'$ is appeased. The core idea of CSDLM is that after the user constructs the BCLCS, the user modifies the original hash pointer information and changes the hash pointer of all data blocks to point to a null address or wrong address. And a new hash value is obtained by calculating the original hash pointer and the $Mask$ to replace the original hash value. The calculated new hash value is the hash pointer in the locked state. The calculation formula is as follows:

$$Hash_{lock} = Hash_{unlock} \oplus Mask \qquad (1)$$

To change from the locked state to the unlocked state, the user only needs to perform an XOR operation on the locked state hash value and the mask value. The calculation is as follows:

$$Hash_{unlock} = Hash_{lock} \oplus Mask \qquad (2)$$

If an attacker obtains a certain data block by illegal means, the attacker cannot obtain other data blocks due to the lack of *Mask* and it is difficult to trace the integral chain.

**4.2.3 Procession of CSDLM**

In the previous chapter, we set up an initial impression of CSDLM. In this chapter, we are going to introduce how dose the CSDLM works in our framework. The procession can be divided into two parts by the dotted line in Fig. 8. The first part is used to encrypt the data, construct BCLCS, transform the hash pointer and store the data. The second part is used to parse the *Meta File*, search data in the decentralized network and recover the data.

**Pre-processing of File.** In this stage, user pre-processing includes five steps:

1. **Parameters generation:** The user first generates the *key* and the *Mask* of the CSDLM. The *key* is a hash value generated according to the mixture of the timestamp and the hash value of the file to be encrypted. Therefore, the key generated each time is random, which improves the security of the data to a certain extent. The key can be expressed as $key = H(timestamp + H(file))$. Similar to the *key*, the *Mask* is also randomly generated by timestamp and so on.

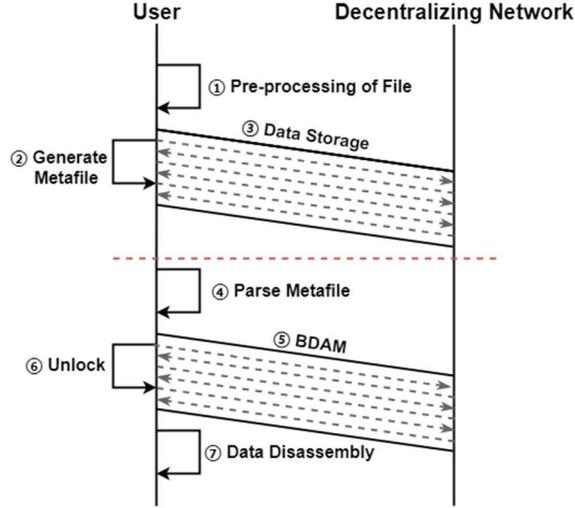

**Fig. 8.** Data Processing Timing Diagram

2. **Data Encryption:** This step is mainly to encrypt the data by *encrypt*(). SM4 symmetric encryption algorithm is adopted in our framework. The encrypted file is represented as: *EF* = *encrypt* (*file*, *key*).

3. **Data Blocking:** This step is mainly for the user to block the ciphertext *EF*. When the ciphertext is divided into $N$ blocks, the ordered set of data blocks can be obtained, which is recorded as: $Blocks = \{Block_i \mid i \in [1, N]\}$.

4. **Key Protection:** Because the security of symmetric cryptographic algorithms often depends on the key, key protection in the file preprocessing stage is very important. Therefore, the framework groups the key and stores it in the block by iteration. The writing location is determined by the *loc*() function which needs to store in the *Meta File*. There is a default mode that does not need to write in *Meta File*, store the subkey at the beginning of the data domain. The relation between the *i*-th subkey and the data block is:

$$key[i] \xrightarrow{write} Block_{(i \bmod N) + 1} \qquad (3)$$

5. **Construction of Chain Structure:** After the key is written, the hash value of the data block is calculated successively, and then the pointer domain of the data block is filled. Finally, the data chain is changed to the locked state through calculation with the *Mask*.

**Generation of Meta File.** The *Meta File* is mainly used to realize the basic mapping between files and nodes. The essential contents and functions of the *Meta File* are shown in Table 5. *Meta File* is usually handed over to users for self-management.

**Table 5.** Basic Elements of the Meta File

| Name | Function |
| --- | --- |
| Head node address ( First Beginner ) | Access to the data chain |
| Header block hash | Access to the correct order of data blocks |
| Mask | Access to the correct data block pointing |

**Data Storage.** When pre-processing is completed, the data storage phase is initiated. The Meta File generation phase coexists with the data storage phase. The period of the generated Meta File ends when the first Beginner node is settled. This phase is stored according to the PoR model. During this phase, P2P transfers are concurrent processing and improve data transfer efficiency.

**4.2.4 Bi-direction Data Access Mechanism (BDAM)**

This section is the second part of the procession of CSDLM. The users obtain the corresponding mapping data from the decentralized network according to the *Meta File* they own. The process is divided into three main stages: explanation of *Meta File* mapping relationships, unlock operation and Bi-direction data acquisition access, and data disassembly.

**Explanation of Meta File Mapping Relationships.** This stage mainly extracts three parameter values in the ordered set block from the *Meta File*, including hash value, Beginner node used to store data block, and *Mask*. The following actions are performed during this phase: The specific operations are as follow:

$$HB_1 = H(Block_1), Beginner, Mask \leftarrow extract(Meta\ File) \qquad (4)$$

**Unlock Operation and Bi-direction Data Acquisition Access.** The *Meta File* is usually used to parse the hash value $HB_1$ of the first data block, and the mapping data pair $\langle HB_1, Beginner \rangle$ of the storage node, which represents the starting point of data acquisition. After requesting the data block with the hash value $HB_1$ from the Beginner, the user restores the pointer domain to the unlocked state in combination with the *Mask*, so as to obtain the correct pointing relationship, namely $Hash_{unlock}$.

In this paper, the way of data acquisition is content addressing [5]. In order to improve the efficiency, we introduce the idea of the Bi-direction replication characteristic of DNA [30] into the **B**i-direction **D**ata **A**cquisition **M**echanism (BDAM). In this mechanism, treating each data block is deoxyribonucleic acid, and accessing the data block Bi-directionally by using a hash pointer to the preceding and following data blocks, thus increasing the speed of data fetching in the decentralized network.

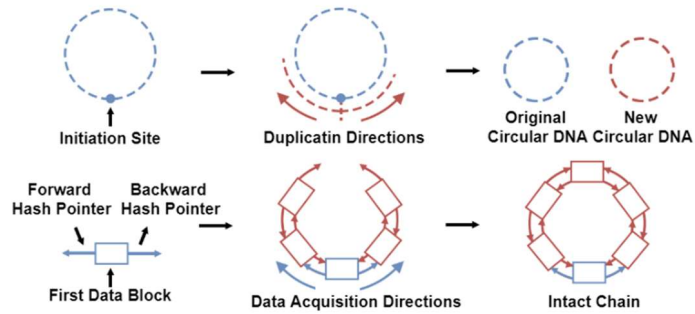

**Fig. 9.** Schematic of Data Acquisition

**Data Disassembly.** This stage can be regarded as the reverse process of the user preprocessing stage. The user first determines the location of the first data block through *Meta File*, parses the key through *loc*() function, then restores the encrypted file according to $EF = \sum_{i=0}^{N} Block_i$, and finally decrypts and restores the original file through *decrypt* (*EF, key*).

### 4.3 The Proof of Resource (PoR) Model

### 4.3.1 Symbol Definition of PoR

The symbols and meanings of PoR are shown in Table 6.

**Table 6.** Symbols of PoR

| Symbol | Description | Symbol | Description |
|---|---|---|---|
| $NF$ | The node file of PoR, records all the node network locations. | $NC$ | The parameter of $Judge()$, presents the capacity of a node. |
| $N$ | Quantity of storage nodes. | $Value$ | The value of a node. |
| $M$ | Quantity of blocks. | $Rate$ | The storage threshold of PoR. |
| $RTT$ | Round-Trip Time. | $freespace$ | The free capacity of a storage node. |
| $Update()$ | Using to update the node file. | $Block_i$ | The $i$-th block. |
| $Nidx$ | The index of a node in the node file. | $CheckRate()$ | Using to justify if a node satisfies the demand of PoR. |
| $random()$ | Using to output an integer. | $Takepart()$ | A campaign event of Follower. |
| $Beginner$ | A node role of PoR. | $Follower$ | A node role of PoR. |
| $Upload()$ | Using to upload a block. | $Judge()$ | Evaluating a node's value. |
| $Election()$ | An election event initiated by Beginner. | $CheckStore()$ | Using to confirm that the data block is saved by the storage node |
| $Sort()$ | The sort function of each node's value. | $k$ | The compress parameter of $Judge()$. |

### 4.3.2 Designation of PoR

**Overview of Model.** In decentralized networks, the differences between networks easily lead to the problem of inconsistency. For network storage, a rapid upload process and sufficient storage space are the expectations of users. However, the traditional decision-making mechanisms sacrifice time to obtain data consistency and have disadvantages such as high requirements for computing power and resource waste. What's more, network circumstances affect the transfer time which will reduce the timeliness of uploading. But they do not consider the impact of the network environment on storage timeliness. In this section, we design a storage decision model based on the node's network environment and storage resources, referred to as PoR.

The model satisfies the following properties:
    1. Low-performance consumption, it does not require high computing power.
    2. Timeliness, it can reach a consensus in time.
    3. Fairness, it can realize fair decision-making, that is, it does not store too many data blocks in a chain at a certain node.

The model must comply with the following guidelines:
    1. All nodes need to jointly maintain and store a node file, denoted as **NF**, recording the address or domain name of the node on the network.
    2. Any node can participate as a data block storage node and wants to participate in storage before resources are exhausted.
    3. For each storage event, the proportion of data blocks stored by a node in the total blocks in a storage event cannot exceed a threshold *Rate*. A provisional records table, which counts the

amounts of blocks that each node has stored for one storage event, is required to secure nodes do not store too many blocks. Once the storage event has been accomplished, the temporary table will be destroyed.

In the PoR model, a storage event contains a user role and two dynamic node roles. The dynamic roles are Beginner and Follower. Beginner owns the store right for the current data block, Follower has no right to store the current data block, but it can interact with Beginner and obtain the store right for the next data block.

Assuming that there are *N* nodes and *M* data blocks in a chain. The steps of selecting storage nodes include preprocessing stage, storage right election stage and data storage inspection, and iterative storage stage. The timing diagram is shown in Fig. 10.

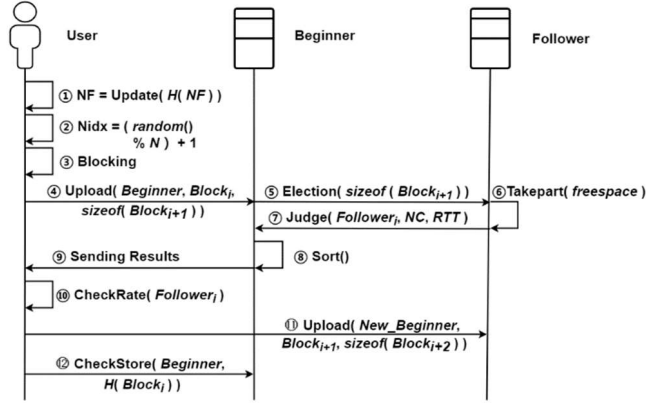

**Fig. 10.** PoR Timing Diagram

**Preprocessing Stage.** This stage mainly completes the selection of the first Beginner. And there are three steps:

1. **Update the Node File:** $NF = Update(H(NF))$, in this step, the user compares the hash value between the local node file *NF* and *NF* in decentralized the network.

2. **Beginner Initialization process:** $Nidx = (random() \% N)+1$, in this step, the user first uses the random function *random*() to obtain the random value and then obtains the serial number *Nidx* of the Beginner through modular calculation. Finally, the *Nidx* node in the node file is selected as the first Beginner.

3. **Storage Event Ready Inform:** $Upload(Beginner, Block_i, sizeof(Block_{i+1}))$, the user informs the Beginner node that it is ready and gives it to the Beginner to store the data block $Block_i$, providing the next block size for the Beginner to initiate an election.

**Storage Campaign Stage.** After the first Beginner is confirmed, this Beginner will initiate the store right of election for the next data block, there are five steps in this stage.

1. **Beginner Initiates Node Storage Campaign:** $Election(sizeof(Block_{i+1}))$, at this stage, Beginner will launch a new data block storage campaign for all Followers.

2. **Follower Campaign Response:** $Takepart(freespace)$, after receiving the campaign notice, the Follower nodes participating in the campaign first judge whether it meets the storage space required by the new node. If its free space satisfies the demands, then respond its free space to the campaign.

3. **Node Evaluating:** $Judge(Follower_i, NC, RTT)$, in this step, Beginner evaluates the *i*-th Follower by its response and **R**ound-**T**rip **T**ime (*RTT*). The evaluation equation needs to meet the following conditions:

    a) The selected node must have good network conditions as far as possible.

    b) The selected node must have as much storage space as possible.

Therefore, the calculation formula of *Value* is as follows:

$$Value = k \cdot \frac{NC}{RTT}, \quad 0<k\leq1, \quad RTT\geq1 \tag{5}$$

Where the parameter *NC* is in Gb and the parameter *RTT* is in ms. If a node refuses to store data or the Beginner doesn't receive the response from the node, then ignore the node this time. The reason why we only choose RTT as the presentative of the network circumstance is that RTT can adequately reflect the real-time situation of the network. If there is a low-quality or busy network, then it has a poor transfer rate and bandwidth. To be quantified, the more depressed RTT will be.

4. **Sorting all the values:** $Sort()$, Beginner sorts all the values in descending order. After that, Beginner sends the result to the user.

5. **User Confirmation:** $CheckRate(Follower_i)$, the user performs a fairness check successively and stops the fairness check until the *i*-th Follower meets the conditions. The check flow is as follows:

    a) Check whether the node stores data blocks. If the node does not store data blocks, select the node.

    b) If all nodes are stored, check whether the *Rate* reaches the threshold.

    c) If all nodes do not meet the requirements, the storage is assigned to the node with the highest Value and the Rate is increased.

**Data Storage Verification and Iterative Storage Stage.** After the completion of a campaign, the user needs to notify the newly selected Beginner and check the validity of the data storage. This stage is divided into three steps.

1. **Newly Storage Event Ready Notification:** The user informs the newly Beginner that ready to store the data block, $Upload(New\_Beginner, Block_{i+1}, sizeof(Block_{i+2}))$.

2. **Storage Verification:** $CheckStore(Beginner, H(Block_i))$, after the Beginner node completes the storage, the user requests the hash value of the data block from the last Beginner, realizes the integrity judgment of the data block through the comparison of the hash value, and prevents the fake storage of the data block.

3. **Loop Processing:** Repeat until all blocks are stored.

It should be stated noted that there is not any incentive to encourage the node to store blocks. The major reasons are as follows. To begin with, we just provide a decentralized storage framework, and the PoR is an indispensable part of it. We do not limit the scope of adoption for this framework, it could not only use on the global internet, but it merely also can use in a local private area network. It depends on the administrator who wants to provide storage services for his groups. In addition, a positive system should constantly improve with its developments. It might be a burden that appends too many affiliations at this beginning phase.

## 5. Framework Analysis

In this section, we present the security analysis and performance analysis of the framework. Firstly,

we prove that our framework is secure. Secondly, it is demonstrated that our BCLCS outperforms blockchain in terms of storage capacity and efficiency performance. Finally, our PoR strategy satisfies the three properties of low-performance consumption, timeliness, and fairness.

### 5.1 Security

### 5.2.1 Symbols Definition of Security Analysis

The symbols used in the analysis and their meanings are shown in Table 7.

**Table 7.** Symbols of Analysis

| Symbol | Description | Symbol | Description |
|---|---|---|---|
| $g()$ | The parameter of the distribution confidentiality. | $\omega()$ | The parameter of the confidentiality of adjacent data blocks. |
| $n_i$ | The storage node. | $b_i$ | The data block of The BCLCS. |
| $C$ | The BCLCS chain. | $\varphi_1$ | The construction law of the BCLCS. |
| $M$ | Quantity of blocks in a chain. | $N$ | Quantity of storage nodes. |
| $a \xrightarrow{\{mess\}} b$ | Send a message to b from a. | $\langle a, b \rangle$ | The relationship between a and b. |
| $e$ | The element of a set. | $S$ | The data set. |
| $f()$ | The function of the CSDLM | $FIN$ | The status of the CSDLM |

### 5.2.2 Security Analysis

In this section, we tend to prove Data distribution confidentiality and Anti-Traverse. These two conceptions make contributions to the framework.

**Distribution confidentiality.** Distribution confidentiality can divide into the confidentiality of adjacent data blocks and the confidentiality of same-node and same-chain blocks.

**Definition 2:** The confidentiality of adjacent data blocks refers to nodes $n_i$ and $n_{i+1}$ can't find out a law $g()$ which can establish the relationship between blocks $b_i$ and $b_{i+1}$. When nodes $n_i$ and $n_{i+1}$ store the block $b_i$ and $b_{i+1}$ respectively.

**Theorem 2:** Our framework satisfies the confidentiality of adjacent data blocks.

**Proof 2:** We assume that there is a law $g()$ and the nodes $n_i$ and $n_{i+1}$ are dishonest, $n_i$ and $n_{i+1}$ will exchange the relationship $\langle b_i, b_{i+1} \rangle$. Then $n_i$ will obtain $\langle b_{i+1}, b_{i+2} \rangle$ through $n_{i+1}$ after $n_{i+1}$ and $n_{i+2}$ exchange $\langle b_{i+1}, b_{i+2} \rangle$. Therefore, any node in the framework will acquire the BCLCS chain $C = \{b_i \mid i \in [1, m]\}$.

However, CSDLM has broken the relationship set up by law $\varphi_1$ before uploading. That causes there isn't relationship $\langle b_i, b_{i+1} \rangle$ which contradicts with assumption. Therefore, the assumption does not hold and it means our framework satisfies the confidentiality of adjacent data blocks.

**Inference 1:** It is impossible that $n_i = n_{i+1}$ and $n_i$ and $n_{i+1}$ store the block $b_i$ and $b_{i+1}$ respectively.

**Proof 3:** For the storage of $b_i$, $user \xrightarrow{\{b_i, sizeof(b_{i+1})\}} n_i$ will occur after determining the $n_i$ to store $b_i$. $n_i$ will campaign $n_i \xrightarrow{\{sizeof(b_{i+1})\}} \{n_j \mid j \in [i, n] \& i \neq j\}$. if $n_i = n_{i+1}$, then the limitation $\{n_j \mid j \in [i, n] \& i \neq j\}$ is unsatisfied. Therefore, the inference is established.

**Definition 3:** the confidentiality of same-node and same-chain blocks is the node $n_i$ can't find out a law $\omega()$ that sets the relationship $\langle b_i, b_j, i \neq j \rangle$. When $n_i$ stores the $b_i$ and $b_j$, $b_i, b_j \in C$.

**Theorem 3:** Our framework satisfies the confidentiality of adjacent data blocks.

**Proof 4:** BCLCS builds the relationship $\langle b_i, b_{i+1} \rangle$ by law $\varphi_1$. But, CSDLM breaks the relationship $\langle b_i, b_{i+1} \rangle$ before uploading. If a node $n_i$ store the $b_i$ and $b_j$, it couldn't find out a low $\omega()$ to set the relationship $\langle b_i, b_j, i \neq j \rangle$. Therefore, this framework processes the confidentiality of same-node and same-chain blocks. Thereby, the $n_i$ can't identify $b_i$ and $b_j$ come from the same BCLCS chain.

**Anti-Traverse.** Anti-Traverse secures the unavailability of a whole BCLCS chain when it is stored in the decentralized network.

**Definition 4:** If a set $S$ is Anti-Traverse, it should exist a status-changing function between two or more statuses. And the status-changing function enables traverse operation in one state but is inaccessible in another state. The traverse operation makes any $e \in S$ accessible.

**Theorem 4:** The data blocks are uploaded to the decentralized network by our framework is Anti-Traverse.

**Proof 5:** If a dishonest role obtains a block $b_i$ by an illegal approach, to recover the whole chain, its next step is to traverse the BCLCS chain through the pointer domain. We assume that the illegal role can traverse the BCLCS chain. Then it can find out the locations where $b_j$, $j \in [1, m]$, and $i \neq j$ are. It collapses with Definition 2 and Definition 3, then the assumption is fake. According to CSDLM, the uploaded blocks are in the $FIN$ status which means it isn't traversable. But it is traversable when the user changes the status of the BCLCS chain by $f()$. Above all, this framework is Anti-Traverse.

## 5.2 Performance Analysis

In this section, we focus on the capacity and efficiency of our framework. Section 5.2.2 will discuss the capacity and section 5.2.3 is efficiency.

### 5.2.1 Symbols Definition of Performance Analysis

The symbols used in the analysis and their meanings are shown in Table 8.

**Table 8.** Symbols of Analysis

| Symbol | Description | Symbol | Description |
| --- | --- | --- | --- |
| $\ll$ | Much smaller than. | $O()$ | The time complexity or space complexity. |
| $n$ | Quantity of storage nodes. | $m$ | Quantity of all blocks. |
| $t$ | The number of calculations for the hash algorithm. | $\delta$ | The consumption of the hash algorithm. |
| $\varepsilon$ | Quantity of blocks in a chain. | $Rate$ | The storage threshold of PoR. |

### 5.2.2 Capacity and Efficiency Analysis

In this chapter, we are going to analyze the storage capacity and efficiency performance. According to the properties of BCLCS, we still presume that there are *n* storage nodes and *m* blocks. The demands of storage capacity for BCLCS is m blocks without redundancy, then the scale is $O(m)$. For redundancy, we design a disaster recovery mechanism in which the storage capacity requirement is triple times of the chain. But it can not only use in this storage system, but it also has a good universality in other circumstances, therefore we would like to present it in our other work. Then the storage capacity is still in the $O(m)$ after redundancy. The searching cost only depends on the number of storage nodes, the escalation of searching is $O(n)$. Compared with blockchain, which is $O(m)$, the escalation of $O(n)$ will have a better performance when the amount of data has explosive growth, which means n ≪ m.

Table 9. Qualitative Comparison of Blockchain and BCLCS

| Object | Capacity Demands | Search Efficiency |
|---|---|---|
| Blockchain | $O(nm)$ | $O(m)$ |
| BCLCS | $O(m)$ | $O(n)$ |

### 5.2.3 The Analysis of PoR

In section 4.3.2, we have required three attributes for PoR, they are low-performance consumption, timeliness, and fairness. In this section, we are going to discuss these three attributes.

**Computing Complexity.** In the computing based mechanism, the major consumption comes from the multiple times' calculations of the hash algorithm. It can be presented as $O(nt\delta)$, t presents the calculation times, $\delta$ denotes the consumption of the hash algorithm. The reason why it multiples a $n$ is that it presents the whole network consumption for a storage event, not a node. In the storage capacity based mechanism, the main consumption comes from the hash calculation after storing some invalid data. The constant-level calculation is only required in this mechanism. Then its computing complexity is $O(n\delta)$. In the PoR, we assume that there are $\varepsilon$ blocks in a chain, $n$ times calculation for equation (5) for each block is demanded. Therefore, $O(\varepsilon n)$ presents its consumption for PoR. $\varepsilon$ is always a small number, so the consumption can transform into $O(n)$.

Table 10. Qualitative Comparison of Consumption

| Type | Computing Based | Capacity Based | PoR |
|---|---|---|---|
| Consumption | $O(nt\delta)$ | $O(n\delta)$ | $O(\varepsilon n)$ or $O(n)$ |

**Timeliness.** In section 4.1, we have stated that blockchain is operating on the second dimension structure. Due to the uncertain creation of blocks, blockchain needs to limit the time gap between neighboring blocks. The primary gap is 10 mins in the blockchain, Ethereum needs about 10 to 12 scends. In PoR, the decision time is reduced to 62.5ms in our case, more detail will argue in section 6.2.2.

**Fairness.** Thanks to the threshold *Rate*, PoR positively limits the number of blocks for a node when it tries to store more blocks in a BSLCS. In the blockchain, most of the computing power is in the hands of a few large enterprises. Although, it is designed to be fair but runs to contrast in practicing.

## 6. Experiments

### 6.1 Experiment Environment

We choose a teaching computer room that has 47 hosts with the same configuration to complete the experiment. All hosts are running the service program of the storage node, and one host is also running the user's program. The configurations of the host are shown in Table 11. The computation-intensive parts such as encryption and decryption are implemented in C in this paper, while the network communication part is completed in Python. The interaction between the two parts is completed through the generated dynamic link library.

**Table 11.** Parameters of Experimental Hosts

| Parameter name | Value |
| --- | --- |
| Central processing unit | Intel(R) Core(TM) i5-10505 CPU @ 3.20GHz     3.19 GHz |
| RAM | 8.00 GB |
| Operating system bits | 64-bit operating system, x64-based processors |
| Operating system | Windows 10 Professional |
| Disk space available | 721GB |

### 6.2 Experiment Results and Analysis

### 6.2.1 Tests and Analysis of Data Transfer Rate

In the experiment, we first set the PoR threshold *Rate* and the data block quantity $N$ to 0.1 and 20 respectively, and then conducted upload tests on files with sizes of 0.5-100.5M. The test results are shown in Fig. 11. The experimental results show that the upload time increases with the increase in file size, and the proportion of encryption time gradually increases, while the proportion of transmission time gradually decreases. Finally, the two tend to a saturation state, which means approaching a stable stage. The total processing time for 100Mb data is 54.18 seconds, and the data processing rate is about 1890Kb/s, which is significantly improved in storage timeliness compared with 95 seconds for processing 20Kb data [16]. Fig. 12. shows the processing rate of the data, in which the curve shows a general upward trend, but the curvature is gradually decreasing.

Different block amounts of $N$ will lead to different saturation values. After making the value of $N$ range from 2 to 14, while keeping the other parameters constant, the data processing rate is measured as shown in Fig. 13. It can be concluded that $N$ is inversely proportionate to the data processing rate. Fig. 13. indicates that with decreasing $N$, the rate of data processing becomes more impacted by the network environment. In other words, the network environment is an important factor in data storage efficiency.

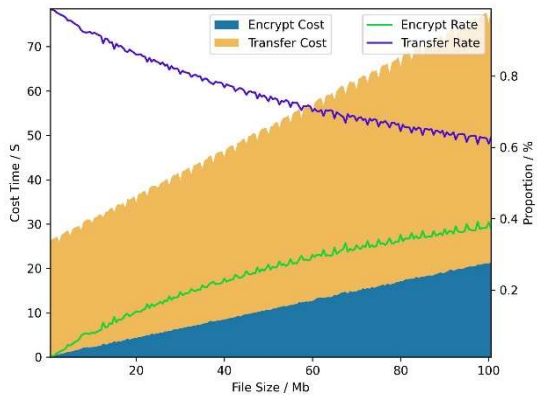
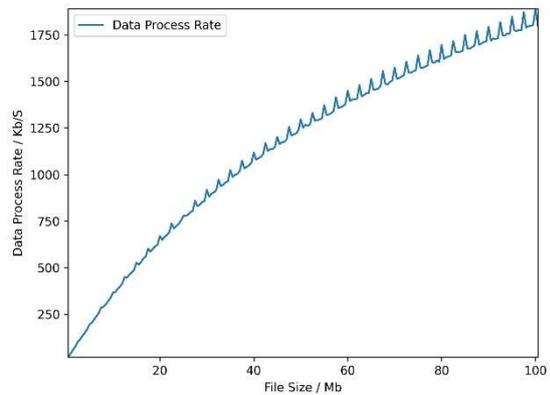

**Fig. 11.** Data Processing Time              **Fig. 12.** Data Processing Rate

The average data processing rate was obtained by calculating different block quantities $N$, and the result is shown in Fig. 14. It is obvious that the average data processing rate decreases smoothly with the increase of block quantity *N*. Combined with Fig. 12. and Fig. 13, different *N* is still consistent with Fig. 12, and the data processing rate eventually tends to be stable.

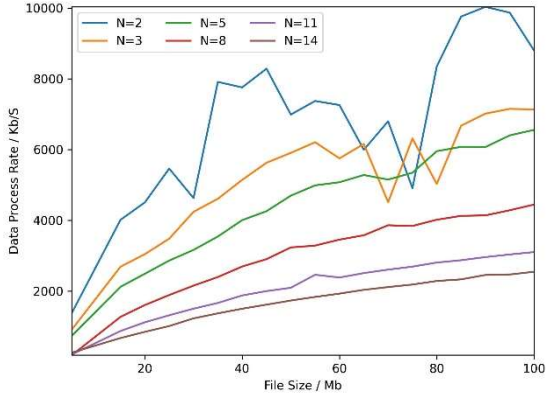 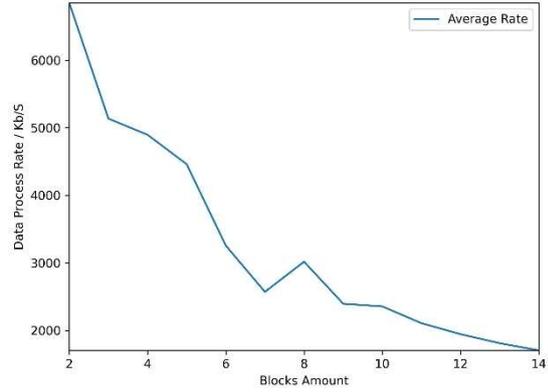

**Fig. 13.** Different N's Data Processing Rate    **Fig. 14.** Average Processing Rate

Based on the analysis of the above experimental results, it is recommended that $N \leq 5$ when the file size is less than 20Mb or the transmission time requirement is high. If the requirements for timeliness are not strict but file security is more important, *N* is recommended to choose $N > 5$.

**6.2.2 Test and Analysis of PoR Decision-making**

In the PoR decision model test, the experimental parameter settings are the same as in Section 5.2.1. We test and analyze the data flow of 4454 data block transfers, and the results are shown in Fig. 15. In the figure, the data block storage of nodes 20 to 47 is significantly lower than that of nodes 1 to 19, indicating that the selection of the first Beginner is random. Due to the same environment of 47 hosts, the value of each node is similar, which leads to the concentration of storage of the first 19 nodes. The experimental results fully reflect the fairness of the PoR decision.

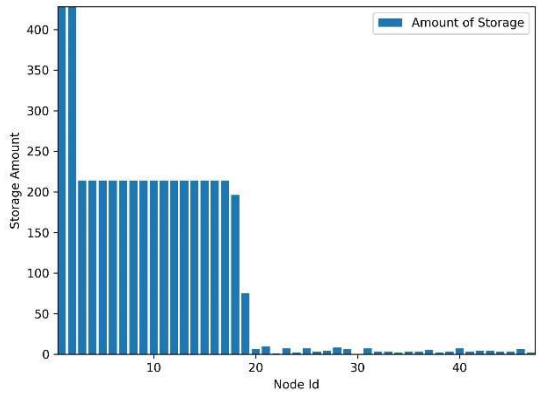 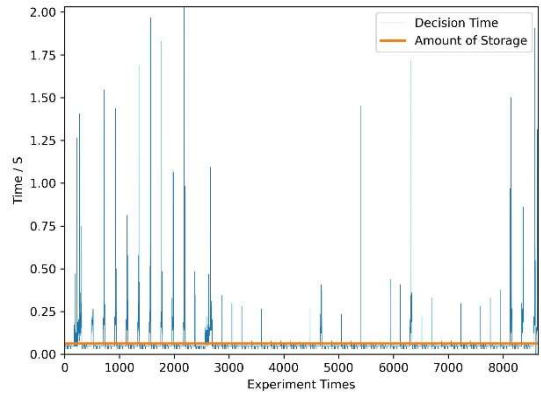

**Fig. 15.** Equivalent Condition Data Block Distribution    **Fig. 16.** PoR Decision Time Diagram

Then we completed 8642 decision monitoring for 47 hosts, and the average decision time of the PoR was calculated to be 62.536 ms. The experimental results are shown in Fig 16. Most of the monitors in the figure are stable below average, with a few exceeding 0.5 seconds, but very few exceeding 1.5 seconds. Due to the sudden network instability between storage nodes, the individual decision time is quite different from the average value. Therefore, for decentralized storage, the impact of the network on storage events must be taken into account. The PoR adds network environment into node evaluation criteria to make up for the lack of network environment consideration.

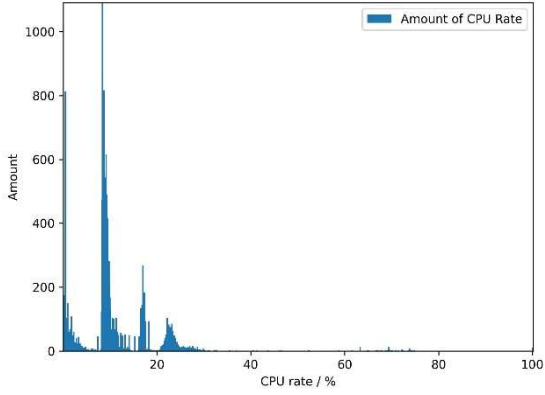 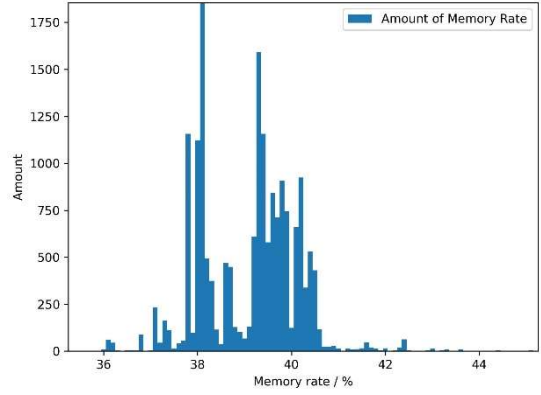

**Fig. 17.** CPU Total Occupancy    **Fig. 18.** Total Memory Occupancy

At the same time, we conducted 18,138 automatic monitoring of host status, and the results are shown in Fig. 17. and Fig. 18, which respectively represent each CPU and memory usage during the detection process. During the experiment, we selected 10 hosts to complete the 10 minutes of idle monitoring, and calculated the average CPU usage as 13.846%, and the average memory usage as 35.656%. By comparison, it can be seen that the PoR decision has negligible influence on the CPU and no obvious influence on the host.

**6.2.3 Experiment and Analysis of Bi-direction Data Acquisition**

In this section, unidirectional and bidirectional data acquisition tests are carried out for data ranging from 0.5Mb to 100.5Mb. Fig. 19. shows the time detection of the two data acquisition methods, and Fig. 20. shows the data processing rate obtained through calculation.

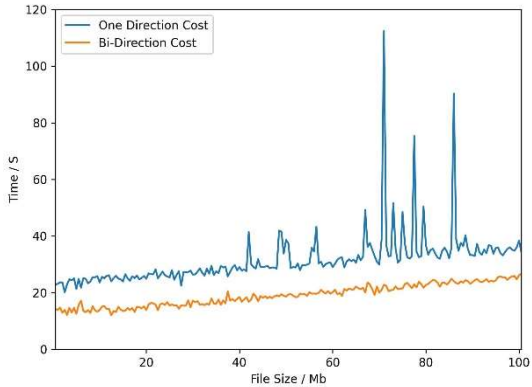 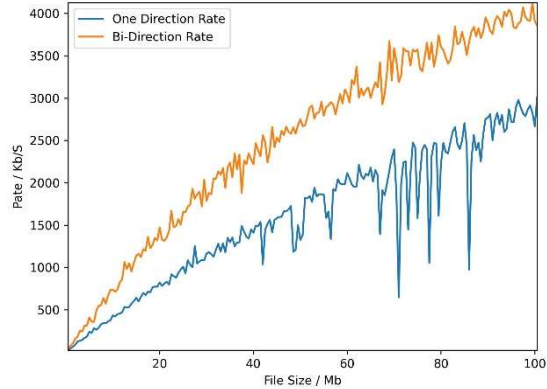

**Fig. 19.** Comparison of Time-consuming for Two Types    **Fig. 20.** Comparison of Processing Rate for Two Types

To better analyze the experimental results, we assume $f(x)$ to be the changing rate of data processing time consumption under the current environment, $DAT_{Bi}(x)$ is the bi-directional data access time calculation function and $DAT_{normal}(x)$ is the single-directional data access time calculation function. The data processing function $DAT$ is determined by the cumulative sum of three time-consuming calculation functions $t_i(x)$, which calculates the data transfer time, the decryption time, and the other operation time. Among which x is the size of the file. The calculation results are shown in Fig. 21. and Table 12. respectively. In this experiment, the BDAM improves the data processing rate by 38.243% on average.

$$f(x) = 1 - \frac{DAT_{Bi}(x)}{DAT_{normal}(x)} \qquad (6)$$

$$DAT(x) = \sum t_i(x), i \in [transform, decrypt, other] \qquad (7)$$

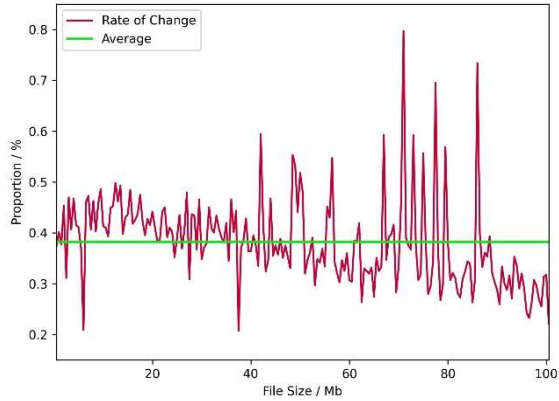

**Fig. 21.** The Data Processing Rate of Change

The above results demonstrate that the use of BDAM effectively reduces data access time, improves data access efficiency, and is more suitable for data storage that requires high timeliness.

**Table 12.** Rate of the Change Data Table

| File Size(Mb) | 10 | 20 | 30 | 40 | 50 |
|---|---|---|---|---|---|
| Rate of Change（%） | 41.34 | 44.17 | 34.79 | 36.42 | 51.82 |
| File Size(Mb) | 60 | 70 | 80 | 90 | 100 |
| Rate of Change（%） | 30.64 | 33.28 | 38.27 | 28.81 | 31.75 |

## 7. Conclusion

In order to solve the problem of decentralized data security storage in the era of big data, a data security storage framework based on BCLCS is proposed. The framework not only improves the storage capacity but also effectively ensures the security of data. Aiming at realizing a more fair and efficient storage right decision-making mechanism in decentralized storage, the paper proposes the PoR, which adds the differences between the networks to the evaluation criteria. This mechanism has low computing power, which greatly improves the timeliness of decision-making on storage rights and ensures the fairness of decision-making. To secure the safety of data in this framework, we combine the SM series algorithms to encrypt and protect the data, and then design the CSDLM to achieve data security storage and access control. Finally, we introduce the idea of DNA Bi-directional replication into BDAM, which significantly improves the data access efficiency. Therefore, the framework proposed in this paper can effectively solve the security problem on the traditional blockchain, and provides a certain research basis for the secure storage of decentralized data.

## 8 Declarations

### 8.1 Ethics Approval

This work does not involve any work related to ethics.

### 8.2 Conflict of Interest

The authors have no relevant financial or non-financial interests to disclose.

### 8.3 Data Availability

The data that support the findings of this study are available from the corresponding author upon reasonable request.

The experimental code for this paper can be accessed at https://github.com/Zijian-Zhou/Haina_Storage_Exp

The engineering prototype developed for this paper is accessible at https://github.com/Zijian-Zhou/Haina_Storage

### 8.4 Funding

The research is supported by the Major Science and Technology Projects of Anhui Province under Grant [No. 201903a05020011]; the Talent Research Fund Project of Hefei University [No. 21-22RC19]; and the University Natural Sciences Research Project of Anhui Province [KJ2021ZD0118]; the National Natural Science Foundation of China Project (62172441, 62172449); the Joint Funds for Railway Fundamental Research of National Natural Science Foundation of China (U2368201); special fund of National Key Laboratory of Ni&Co Associated Minerals Resources Development and Comprehensive Utilization (GZSYS-KY-2022-018, GZSYS-KY-2022-024); Key Project of Shenzhen City Special Fund for Fundamental Research (JCYJ20220818103200002); the National Natural Science Foundation of Hunan Province (2023JJ30696).

### 8.5 Consent to publish

All authors reviewed and approved to submit the manuscript.